\documentclass{PoS}
\usepackage{hepunits}
\usepackage{subfigure}
\usepackage{color}
\usepackage{url}
\usepackage{soul}

\newcommand{\tanb}{\tan\beta}
\newcommand{\sgnmu}{\mathrm{sgn}(\mu)}
\newcommand{\fittino}{\textsc{Fittino}}
\newcommand{\HS}{\textsc{HiggsSignals}}

\newcommand{\ifb}{\mathrm{fb}^{-1}}

\newcommand{\sstop}{\tilde{t}}

\ReportNo{BONN-TH-2013-19}
\title{Constrained Supersymmetry after the Higgs Boson Discovery: A global analysis with \fittino}

\ShortTitle{Constrained SUSY after the Higgs Boson Discovery}

\author{Philip Bechtle, Klaus Desch, Bj\"orn Sarrazin, Mathias Uhlenbrock, Peter Wienemann \\
        Physikalisches Institut, Bonn University, Germany\\
        E-mail: \email{bechtle@physik.uni-bonn.de}, \email{desch@physik.uni-bonn.de}, \email{sarrazin@physik.uni-bonn.de}, \email{uhlenbrock@physik.uni-bonn.de}, \email{wienemann@physik.uni-bonn.de}}

\author{Herbert K. Dreiner, \speaker{Tim Stefaniak} \\ 
        Physikalisches Institut and Bethe Center for Theoretical Physics, Bonn University, Germany\\
        E-mail: \email{dreiner@th.physik.uni-bonn.de}, \email{tim@th.physik.uni-bonn.de}}

\author{Matthias Hamer \\
	II. Physikalisches Institut, University of G\"ottingen, G\"ottingen, Germany\\
	E-mail: \email{mhamer@uni-goettingen.de}}
	
\author{Michael Kr\"amer \\
        Institute for Theoretical Particle Physics and Cosmology, RWTH Aachen, Germany\\
        E-mail: \email{mkraemer@physik.rwth-aachen.de}}

\author{Werner Porod, Ben O'Leary \\
        Institut f\"ur Theoretische Physik und Astrophysik, University of W\"urzburg, Germany\\
        E-mail: \email{porod@physik.uni-wuerzburg.de}, \email{ben.oleary@physik.uni-wuerzburg.de}}

\author{Xavier Prudent \\
	Institut f\"ur Kern- und Teilchenphysik, TU Dresden, Dresden, Germany\\
        E-mail: \email{prudent@physik.tu-dresden.de}}

\abstract{
We present preliminary results from the latest global fit analysis of the constrained minimal supersymmetric standard model (CMSSM) performed within the \fittino\ framework. The fit includes low-energy and astrophysical observables as well as collider constraints from the non-observation of new physics in supersymmetric searches at the LHC. Furthermore, the Higgs boson mass and signal rate measurements from both the LHC and Tevatron experiments are included via the program \HS. Although the LHC exclusion limits and the Higgs mass measurements put tight constraints on the viable parameter space, we find an acceptable fit quality once the Higgs signal rates are included. 
}

\FullConference{The European Physical Society Conference on High Energy Physics -EPS-HEP2013\\
		18-24 July 2013\\
		Stockholm, Sweden}

\begin{document}

\section{Introduction}

Supersymmetry (SUSY) provides an elegant solution to the fine tuning problem of the Standard Model (SM) if supersymmetric particles are realized around the $\TeV$ scale. Then, the SUSY parameter space underlies various constraints from SM and astrophysical observables as well as direct sparticle searches at colliders. Moreover, since the Higgs boson masses and couplings are predictions of the theory, the mass and signal strength measurements of the recently discovered Higgs boson~\cite{Aad:2012tfa} severely constrain the parameter space. We perform a global fit of the constrained Minimal Supersymmetric Standard Model (CMSSM) to these observables in order to answer the following questions: (\textit{i}) What is the allowed SUSY model parameter space after including all available and relevant observables and constraints? (\textit{ii}) To what extend are the observables and constraints in mutual agreement?


The CMSSM is defined by a few parameters at the grand unification (GUT) scale $\sim 10^{16}~\GeV$: Universal soft-breaking mass parameters for the scalars and gauge fermions, $M_0$ and $M_{1/2}$, respectively, a universal soft-breaking trilinear coupling, $A_0$, the ratio of the vevs of the two Higgs doublets, $\tanb$, and the sign of the Higgs mixing parameter, $\sgnmu$, which we fix to be positive in our study. In addition to the four free CMSSM fit parameters, we allow for the top quark mass $m_t$ as an additional nuisance parameter.

In a previous \fittino\ analysis~\cite{Bechtle:2012zk} we found rather grim prospects for a discovery of supersymmetric effects within the CMSSM for the near future: The sparticles and the remaining Higgs spectrum was most likely beyond the LHC $8~\TeV$ reach. The light Higgs boson signal rates and the branching ratio for $B_s \to \mu\mu$ were predicted to be close to their SM prediction. Furthermore, no dark matter (DM) signal was expected in current direct and indirect searches.
This presentation updates and extends the previous analysis~\cite{Bechtle:2012zk} by an implementation of the Higgs boson signal rate measurements. Here, we show first preliminary results of the on-going work, which will be presented in more detail in a future publication.

\section{The \fittino\ framework and technical implementation}

The \fittino\ framework~\cite{Bechtle:2004pc,Bechtle:2009ty} incorporates an auto-adaptive Markov Chain Monte Carlo (MCMC) algorithm, allowing to efficiently sample the multi-dimensional SUSY parameter space. The results presented here are based on a high statistics sample with $\sim 10^8$ scan points and a purely frequentist interpretation. For each scan point, the SUSY particle spectrum is calculated with \textsc{SPheno-3.1.11}~\cite{Porod:2003um}, followed by a dedicated evaluation of the Higgs masses and couplings as well as the SUSY contribution to the anomalous magnetic moment of the muon, $a_\mu^\mathrm{SUSY}$, with \textsc{FeynHiggs-2.9.4}~\cite{Heinemeyer:1998yj}. We use \textsc{SuperIso-3.3}~\cite{Mahmoudi:2007vz} for predictions of the heavy flavor observables and \textsc{MicrOMEGAs-2.4.5}~\cite{Belanger:2001fz} for the DM relic density calculation. \textsc{AstroFit}~\cite{Nguyen:2012rx} and \textsc{DarkSUSY-5.0.5}~\cite{Gondolo:2004sc} are employed to include direct detection limits from DM searches. We include exclusion limits from Higgs boson searches at LEP, Tevatron and LHC using \textsc{HiggsBounds-3.8.1}~\cite{Bechtle:2008jh,Bechtle:2013gu}. The LEP constraints on Higgs boson production are incorporated as a reconstructed $\chi^2$ likelihood~\cite{Bechtle:2012zk,Bechtle:2013gu}. The $\chi^2$ contribution from the Higgs boson signal rate and mass measurements from the Tevatron and LHC experiments is evaluated with \textsc{HiggsSignals-1.0.0}~\cite{Bechtle:2013xfa}.

\section{Experimental constraints}

A detailed description of the experimental measurements and constraints (including references) can be found in Ref.~\cite{Bechtle:2012zk}. Here, we briefly discuss the new or updated observables in the current fit and their implementation.

We include the LHCb measurement of $\mathrm{BR}(B_s\to\mu\mu) = (3.20 \pm 1.50 \pm 0.76) \cdot 10^{-9}$~\cite{Aaij:2012nna}, the updated Belle measurement of $\mathrm{BR}(B\to\tau\nu) = (0.72 \pm 0.27 \pm 0.11 \pm 0.07) \cdot 10^{-4}$~\cite{Adachi:2012mm} and the new relic density abundance determination from the Planck collaboration, $\Omega_\mathrm{CDM} h^2 = 0.1187 \pm 0.0017 \pm 0.0119$~\cite{Ade:2013zuv}, in our fit. 
Note, that the updated $\mathrm{BR}(B\to\tau\nu)$ measurement is in better agreement with both the SM and CMSSM predictions than the old value used in Ref.~\cite{Bechtle:2012zk}.

\begin{figure}
\centering
\subfigure[Comparison of our reconstructed likelihood with the ATLAS $95\%$ C.L. exclusion contours.]{\includegraphics[width=0.43\textwidth]{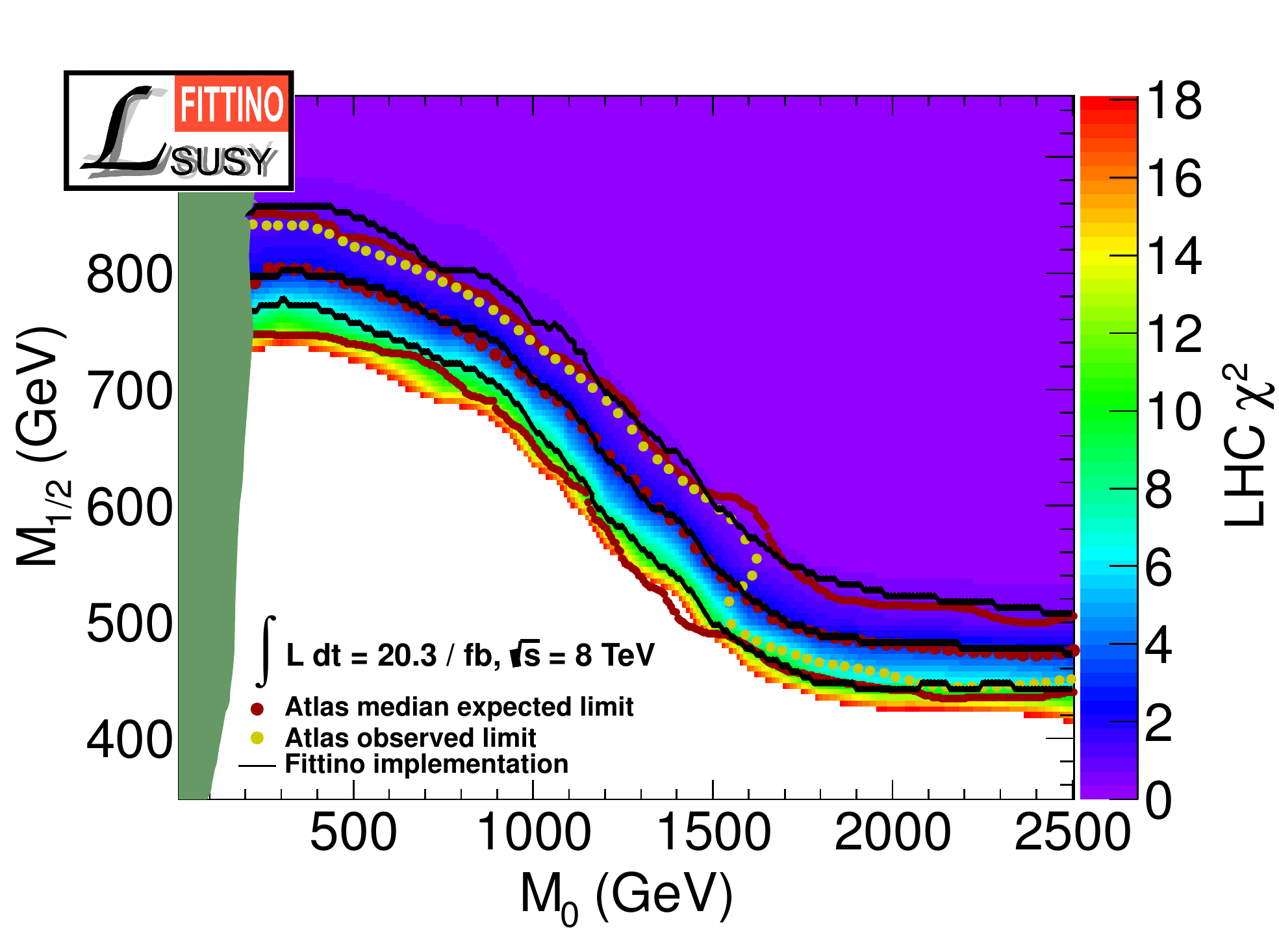}}
\hspace{0.4cm}
\subfigure[Relative difference of the grid-interpolated $\chi^2$ and the true $\chi^2$ contribution 
in the ($A_0, \tanb$) plane, examplarily shown for $(M_0,~M_{1/2})=(1400,~495)~\GeV$.]{\includegraphics[width=0.47\textwidth]{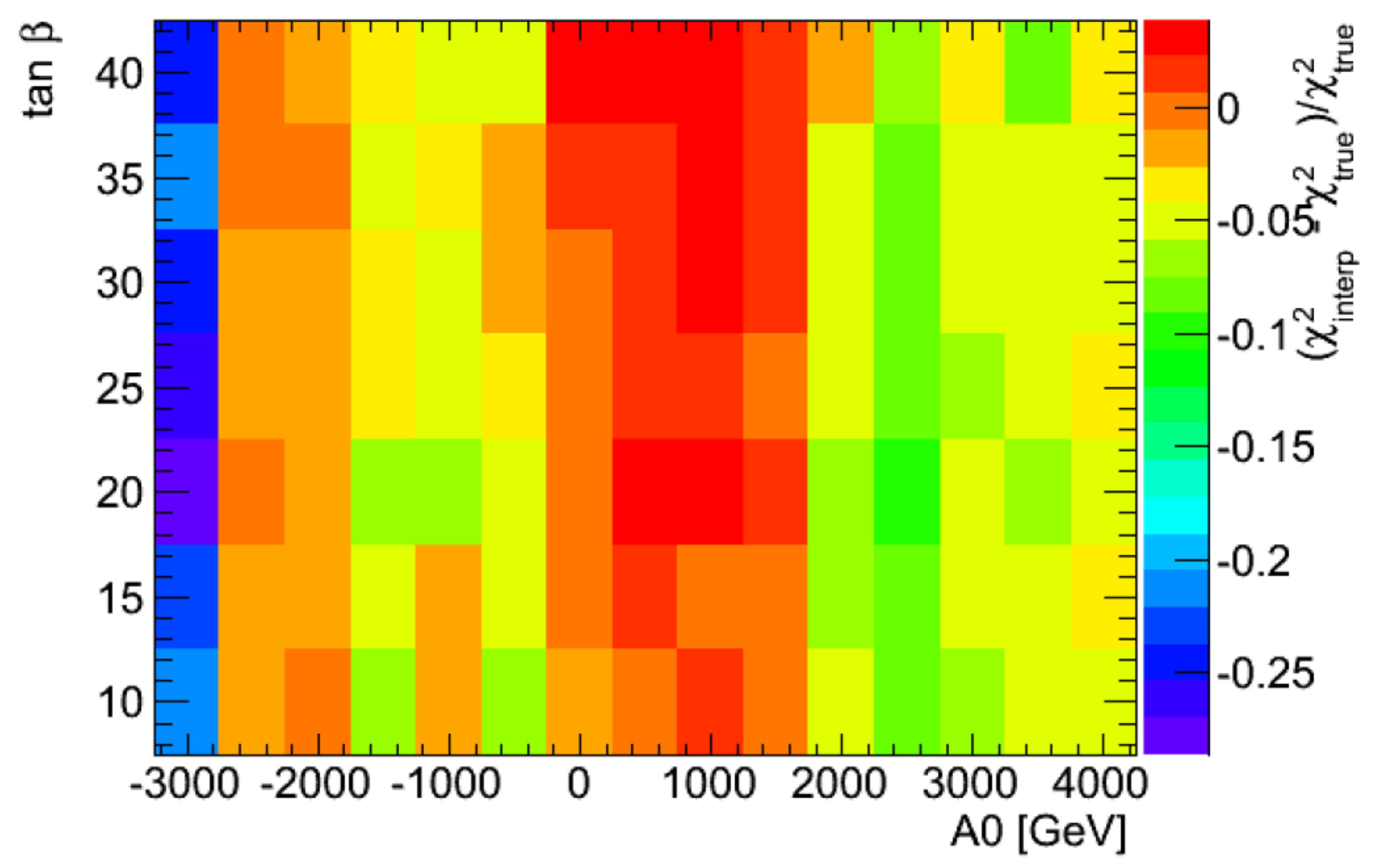}}
\caption{Implementation of the exclusion limit from the full hadronic ATLAS SUSY search~\cite{ATLAS-CONF-2013-047}.}
\label{Fig:LHC}
\end{figure}

We update our implementation of the LHC constraints from searches for direct sparticle pair production (see Ref.~\cite{Bechtle:2012zk} for a description) to accommodate the latest ATLAS results in the full hadronic channel using $20.3~\ifb$ of data at $\sqrt{s}=8~\TeV$~\cite{ATLAS-CONF-2013-047}. In Fig.~\ref{Fig:LHC}(a) we show a comparison of the reproduced likelihood with the official ATLAS exclusion limits in the $(M_0,~M_{1/2})$ plane. We find that in the parameter region around the exclusion limit, the contribution from light $\sstop_1$ pairs to the signal yield is non-negligible.
This introduces a dependence 
 on the trilinear soft-breaking parameter $A_0$, which strongly influences the $\sstop_1$ mass in the renormalization group (RG) evolution. We found that the uncorrected $\chi^2$ contribution (using only the $(M_0,~M_{1/2})$ acceptance grid) can be smaller than the true $\chi^2$ contribution (obtained from full Monte-Carlo simulation) by up to $80\%$ for $|A_0| \gtrsim 2-3~\TeV$. Therefore, we construct additional acceptance grids in the $(A_0,~\tanb)$ plane for $(M_0,~M_{1/2})$ values along the exclusion contour. After this correction, the differences between the acceptance grid based $\chi^2$ and the true $\chi^2$ are typically $\lesssim 10\%$, see Fig.~\ref{Fig:LHC}(b), which can be considered as the systematic uncertainty of our implementation.

We include the 47 Higgs boson signal rate measurements from the Tevatron and LHC experiments provided by \textsc{HiggsSignals-1.0.0} (see Ref.~\cite{Bechtle:2013xfa} for a description and references). Moreover, we include the four Higgs mass measurements from the ATLAS and CMS $H\to \gamma\gamma$ and $H\to ZZ^{(*)}\to 4\ell$ analyses. Note, that \textsc{HiggsSignals} treats the uncertainties for the signal rate and Higgs mass predictions as well as the luminosity uncertainty as fully correlated Gaussian uncertainties. We use the same estimates for the Higgs boson production cross section and branching ratio uncertainties for the SM and the CMSSM and assume a theoretical (Gaussian) uncertainty of $3~\GeV$ on the light Higgs boson mass prediction.

\section{Results}

\begin{figure}
\centering
\subfigure[($M_0,~M_{1/2}$) plane.]{\includegraphics[width=0.48\textwidth]{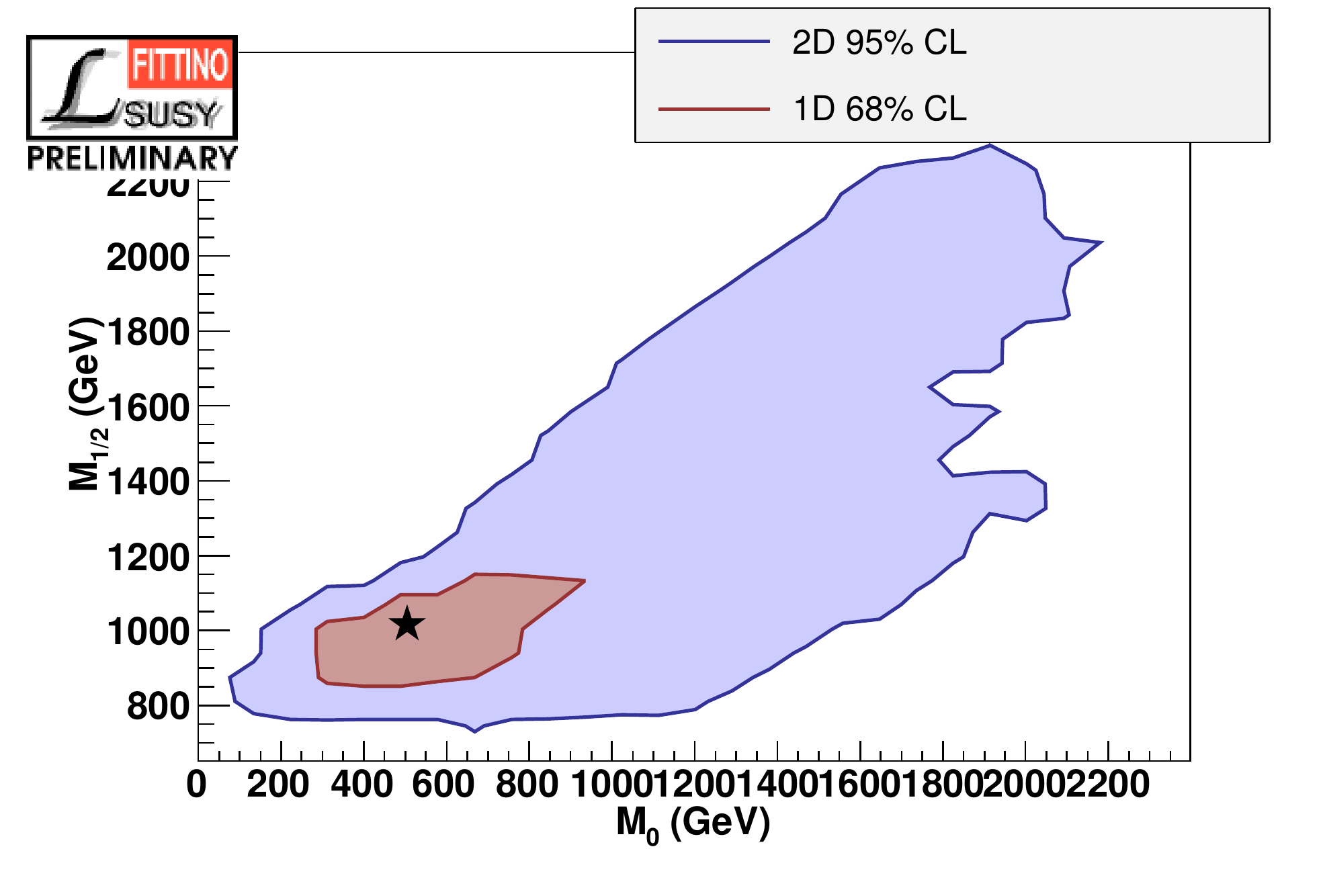}}
\subfigure[($A_0,~\tanb$) plane.]{\includegraphics[width=0.48\textwidth]{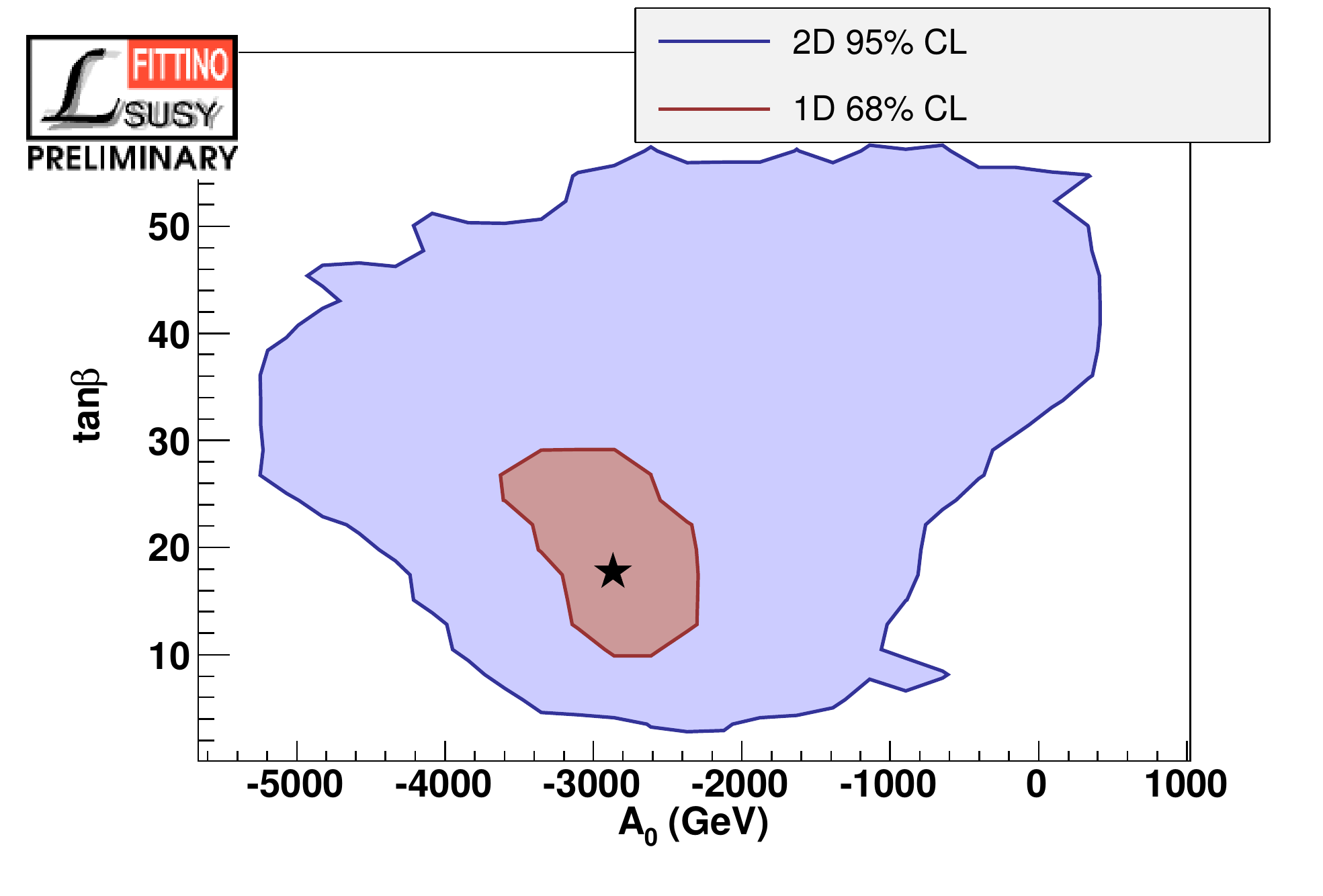}}
\caption{Preferred two-dimensional CMSSM parameter regions.
We profile over the remaining fit parameters. The red and blue areas correspond to the one-dimensional $68\%$ C.L. ($\Delta\chi^2 = 1$) and two-dimensional $95\%$ C.L. ($\Delta\chi^2 = 6.18$) parameter region, respectively. The black star indicates the best-fit point.}
\label{Fig:BFregions}
\end{figure}

We show the preferred parameter space in the $(M_0,~M_{1/2})$ and $(A_0,~\tanb)$ plane in Fig.~\ref{Fig:BFregions}. After including the Higgs boson observables, we find that the stau co-annihilation region is preferred over the focus-point region since it can accommodate slightly better the correct Higgs boson mass. Governed by the observed Higgs boson mass, the fit prefers large negative values of $A_0$, where the stop mixing is (nearly) maximized, and a vanishing trilinear coupling, $A_0 = 0~\GeV$, is disfavored. The parameter $\tanb$ is rather unconstrained. We find a weak linear correlation between $\tanb$ and $M_0$ for the preferred parameter space. 
The best-fit point is found at
\begin{equation}
M_0 = 504~\GeV,~M_{1/2}=1016~\GeV,~A_0=-2870~\GeV,~\tanb=18,~m_t = 173.74~\GeV,
\end{equation}
with a good fit quality of $\chi^2/\mathrm{ndf} = 49.6/59$. The tension observed in previous studies~\cite{Bechtle:2012zk} between the direct LHC limits and the Higgs boson mass measurements (preferring a heavy colored SUSY particle spectrum) versus the anomalous magnetic moment of the muon (preferring a light uncolored SUSY particle spectrum) becomes more severe with the updated LHC exclusion limit. In the remaining viable CMSSM parameter space the SUSY contribution to the anomalous magnetic moment is rather negligible. This tension motivates the consideration of more general models which abrogate the strong connection of the colored and uncolored sparticle masses of the CMSSM.

With the inclusion of the Higgs boson signal rate measurements --- which generally are in very good agreement with the predictions for the SM Higgs boson --- the fit quality improves significantly. This is because the CMSSM parameter regions, which are not excluded by the LHC SUSY searches, naturally feature a decoupled heavy Higgs spectrum and therefore a SM-like lightest Higgs boson.

In Fig.~\ref{Fig:allowedratesandmasses}(a) we show the (SM normalized) Higgs boson partial decay widths, normalized to the $h\to ZZ$ decay mode, for the preferred parameter regions. Deviations from the SM prediction are at most $\sim \mathcal{O}(3\%)$, making the CMSSM extremely difficult to probe via Higgs boson signal rate measurements even at a future linear collider. Note also, that there are remaining theoretical uncertainties of the \textsc{FeynHiggs} calculation of these rates.

The sparticle and Higgs boson mass spectrum predicted by the fit is shown in Fig.~\ref{Fig:allowedratesandmasses}(b). The heavy Higgs bosons with masses $\gtrsim1~\TeV$ are beyond the LHC reach. The typical squark mass scale is $\sim 2~\TeV$, only the $\sstop_1$ may be as light as $\gtrsim 750~\GeV$. The lightest neutralino and stau are nearly mass degenerate with masses around $350 - 600~\GeV$.


\begin{figure}
\centering
\subfigure[(SM normalized) partial decay widths of the lightest Higgs boson, normalized to the $h\to ZZ$ decay mode.]{\includegraphics[width=0.46\textwidth]{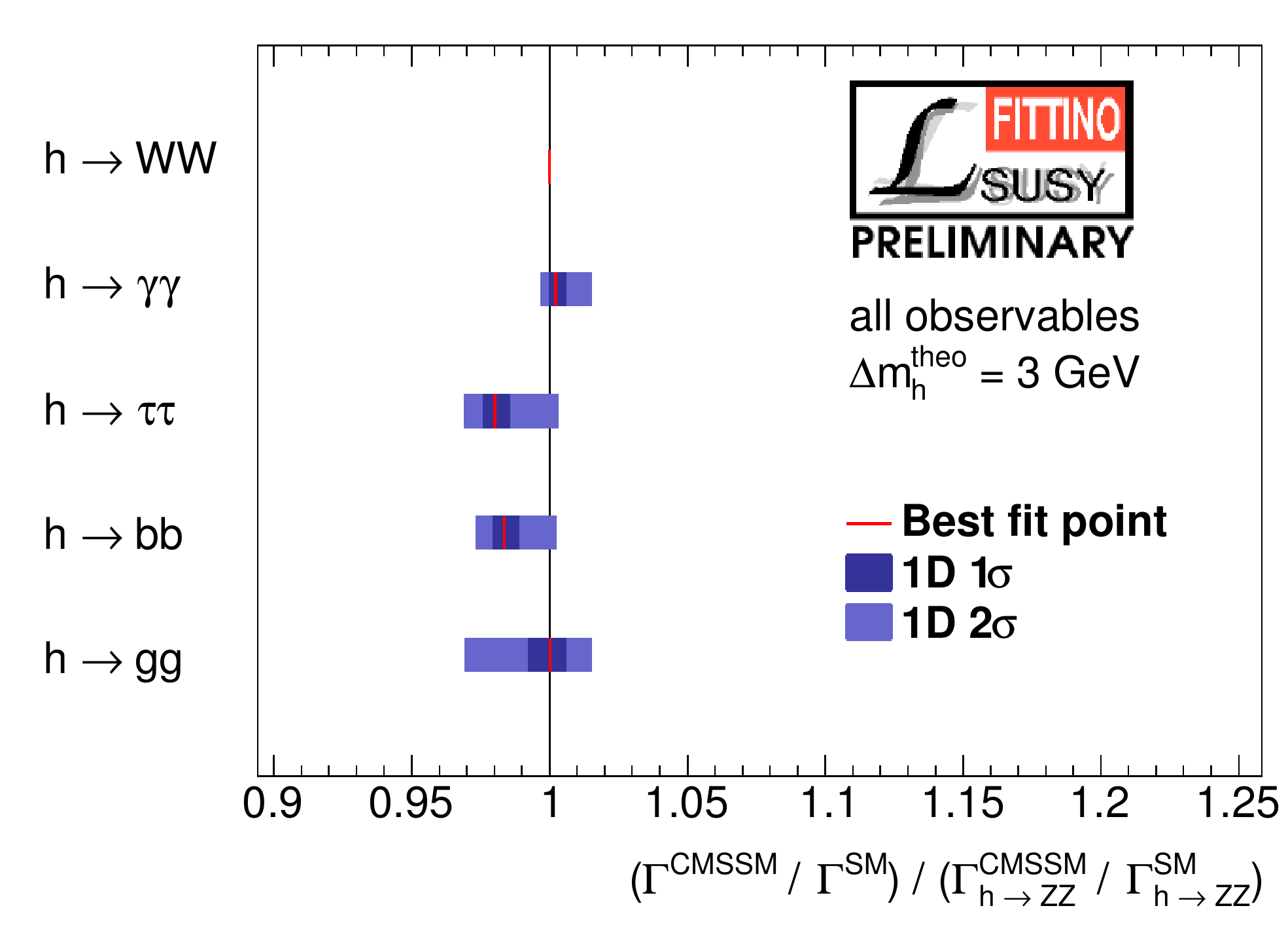}}
\subfigure[Mass spectrum of the Higgs bosons and sparticles.]{\includegraphics[width=0.46\textwidth]{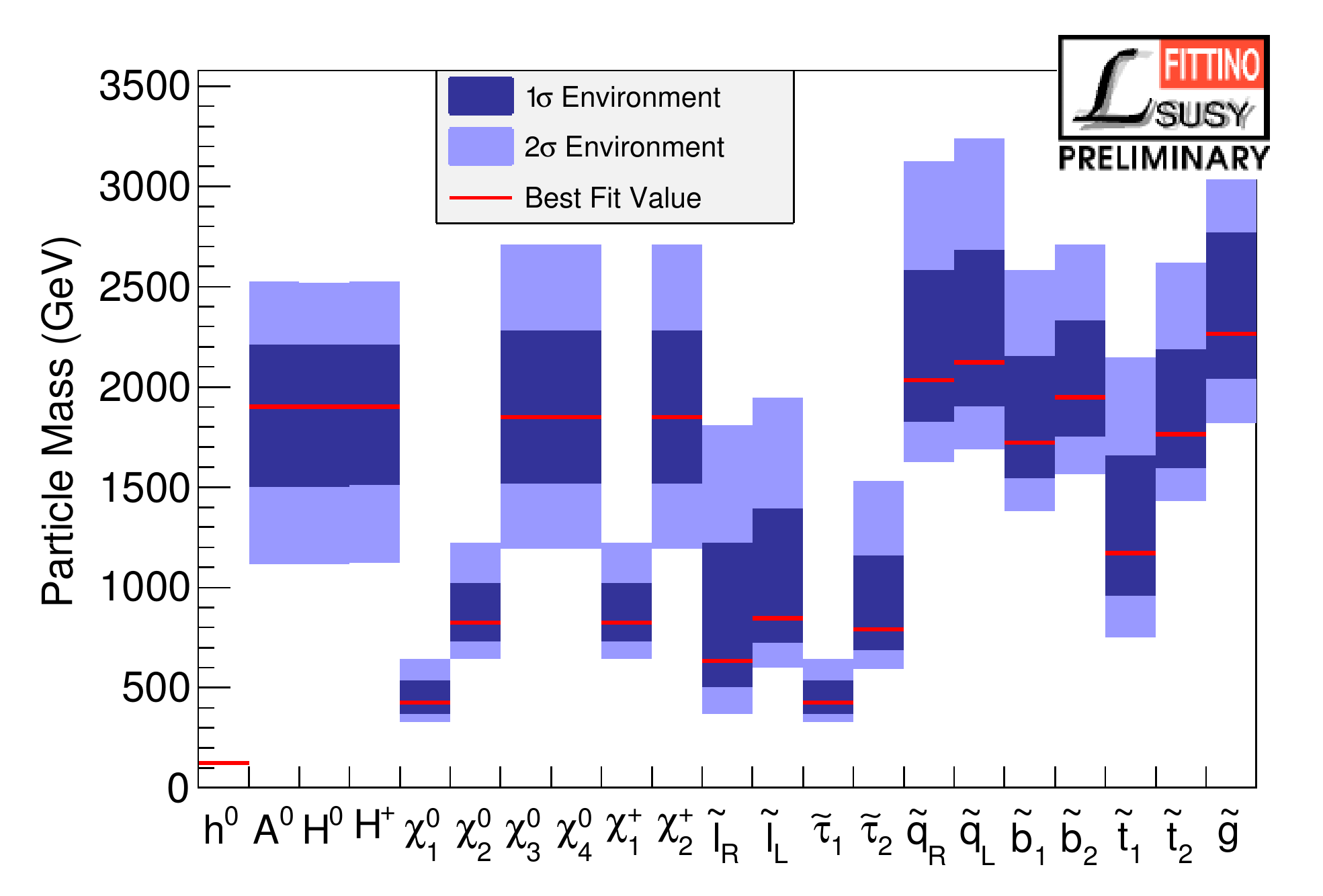}}
\caption{Predictions for the Higgs and SUSY particle spectrum in the allowed parameter region.}
\label{Fig:allowedratesandmasses}
\end{figure}

\begin{figure}
\centering

\end{figure}

\section{Conclusions and outlook}

We presented preliminary results from an ongoing \fittino\ global fit analysis of the CMSSM, including various up-to-date observables and constraints from low-energy and flavor physics, astrophysics, direct LHC SUSY searches and Higgs boson searches. For the lightest Higgs boson we include the Higgs mass and signal rate measurements from the Tevatron and LHC experiments using the program \textsc{HiggsSignals}. The direct LHC SUSY limit and the Higgs boson mass measurements drive the fit to regions with very heavy sparticles, making the CMSSM incapable of explaining the discrepancies observed in the anomalous magnetic moment of the muon. The inclusion of the Higgs signal rates improves the fit quality, since the already preferred region naturally features a SM-like light Higgs boson. However, it does not change significantly the general picture. The prospects for a future direct or indirect discovery of the CMSSM are rather grim. Phenomenologically, the CMSSM looks like the SM with a viable explanation of dark matter. 

The next steps within this project comprise a dedicated $p$-value calculation with repeated fits to randomly generated pseudo-measurements in order to provide a quantitative statement about the fit quality. Furthermore, we plan to investigate the fit outcome and $p$-value dependence using a more inclusive Higgs boson signal rate observable set with $\sim \mathcal{O}(10)$ measurements.

\acknowledgments

We thank the organizers of the EPS HEP 2013 conference for the opportunity to present this work. TS is grateful for the warm hospitality and support of \textsc{Nordita} during his extended stay in Stockholm and thanks the participants and organizers of the \textsc{Nordita} Workshop ``Beyond the LHC'' for a stimulating workshop in a very pleasant atmosphere.


\begin{thebibliography}{99}


\bibitem{Aad:2012tfa}
  G.~Aad {\it et al.}  [ATLAS Collaboration],
  Phys.\ Lett.\ B {\bf 716} (2012) 1
  [arXiv:1207.7214 [hep-ex]];
  S.~Chatrchyan {\it et al.}  [CMS Collaboration],
  Phys.\ Lett.\ B {\bf 716} (2012) 30
  [arXiv:1207.7235 [hep-ex]].

\bibitem{Bechtle:2012zk}
  P.~Bechtle, T.~Bringmann, K.~Desch, H.~Dreiner, M.~Hamer, C.~Hensel, M.~Kramer and N.~Nguyen {\it et al.},
  JHEP {\bf 1206} (2012) 098
  [arXiv:1204.4199 [hep-ph]].

\bibitem{Bechtle:2004pc}
  P.~Bechtle, K.~Desch and P.~Wienemann,
  Comput.\ Phys.\ Commun.\  {\bf 174} (2006) 47
  [hep-ph/0412012].

\bibitem{Bechtle:2009ty}
  P.~Bechtle, K.~Desch, M.~Uhlenbrock and P.~Wienemann,
  Eur.\ Phys.\ J.\ C {\bf 66} (2010) 215
  [arXiv:0907.2589 [hep-ph]].

\bibitem{Porod:2003um}
  W.~Porod,
  Comput.\ Phys.\ Commun.\  {\bf 153} (2003) 275
  [hep-ph/0301101];
  W.~Porod and F.~Staub,
  Comput.\ Phys.\ Commun.\  {\bf 183} (2012) 2458
  [arXiv:1104.1573 [hep-ph]].

\bibitem{Heinemeyer:1998yj}
  S.~Heinemeyer, W.~Hollik and G.~Weiglein,
  Comput.\ Phys.\ Commun.\  {\bf 124} (2000) 76
  [hep-ph/9812320];
  S.~Heinemeyer, W.~Hollik and G.~Weiglein,
  Eur.\ Phys.\ J.\ C {\bf 9} (1999) 343
  [hep-ph/9812472];
  G.~Degrassi, S.~Heinemeyer, W.~Hollik, P.~Slavich and G.~Weiglein,
  Eur.\ Phys.\ J.\ C {\bf 28} (2003) 133
  [hep-ph/0212020].
  
\bibitem{Mahmoudi:2007vz}
  F.~Mahmoudi,
  Comput.\ Phys.\ Commun.\  {\bf 178} (2008) 745
  [arXiv:0710.2067 [hep-ph]];
  F.~Mahmoudi,
  Comput.\ Phys.\ Commun.\  {\bf 180} (2009) 1579
  [arXiv:0808.3144 [hep-ph]].
  
\bibitem{Belanger:2001fz}
  G.~Belanger, F.~Boudjema, A.~Pukhov and A.~Semenov,
  Comput.\ Phys.\ Commun.\  {\bf 149} (2002) 103
  [hep-ph/0112278];
  G.~Belanger, F.~Boudjema, A.~Pukhov and A.~Semenov,
  arXiv:1305.0237 [hep-ph].
  
\bibitem{Nguyen:2012rx}
  N.~Nguyen, D.~Horns and T.~Bringmann,
  arXiv:1202.1385 [astro-ph.HE].
\bibitem{Gondolo:2004sc}
  P.~Gondolo, J.~Edsj\"o, P.~Ullio, L.~Bergstr\"om, M.~Schelke and E.~A.~Baltz,
  JCAP {\bf 0407} (2004) 008
  [astro-ph/0406204].
  
  
\bibitem{Bechtle:2008jh}
  P.~Bechtle, O.~Brein, S.~Heinemeyer, G.~Weiglein and K.~E.~Williams,
  Comput.\ Phys.\ Commun.\  {\bf 181} (2010) 138
  [arXiv:0811.4169 [hep-ph]];
  P.~Bechtle, O.~Brein, S.~Heinemeyer, G.~Weiglein and K.~E.~Williams,
  Comput.\ Phys.\ Commun.\  {\bf 182} (2011) 2605
  [arXiv:1102.1898 [hep-ph]].

\bibitem{Bechtle:2013gu}
  P.~Bechtle, O.~Brein, S.~Heinemeyer, O.~St{\aa}l, T.~Stefaniak, G.~Weiglein and K.~Williams,
  PoS CHARGED {\bf 2012} (2012) 024
  [arXiv:1301.2345 [hep-ph]]. See also new manual available on \url{http://higgsbounds.hepforge.org}.
 
\bibitem{Bechtle:2013xfa}
  P.~Bechtle, S.~Heinemeyer, O.~St{\aa}l, T.~Stefaniak and G.~Weiglein,
  arXiv:1305.1933 [hep-ph];
  O.~St{\aa}l and T.~Stefaniak,
  in proceedings of \emph{The European Physical Society Conference on High Energy Physics},
  \pos{PoS(EPS-HEP 2013)314}.
  
\bibitem{Aaij:2012nna}
  R.~Aaij {\it et al.}  [LHCb Collaboration],
  Phys.\ Rev.\ Lett.\  {\bf 110} (2013) 021801
  [arXiv:1211.2674 [hep-ex]].
 
\bibitem{Adachi:2012mm}
  I.~Adachi {\it et al.}  [Belle Collaboration],
  Phys.\ Rev.\ Lett.\  {\bf 110} (2013) 131801
  [arXiv:1208.4678 [hep-ex]].
 
 
\bibitem{Ade:2013zuv}
  P.~A.~R.~Ade {\it et al.}  [Planck Collaboration],
  arXiv:1303.5076 [astro-ph.CO].
  
\bibitem{ATLAS-CONF-2013-047}
  The ATLAS Collaboration,
  ATLAS-CONF-2013-047.

  
\end{thebibliography}
\end{document}